\documentstyle[preprint,aps,epsf]{revtex}

\def\be {\begin{eqnarray}}
\def\ee {\end{eqnarray}}
\def\beq {\begin{equation}}
\def\eeq {\end{equation}}

\def\bi {\begin{itemize}}
\def\ei {\end{itemize}}
\def\ben {\begin{enumerate}}
\def\een {\end{enumerate}}
\def\ni {\noindent}
\def\vpar {v_\parallel}
\def\vperp {v_\perp }
\def\dpar {\Delta p^\parallel }
\def\dperp {\Delta p^\perp }
\def\qpar {q_\parallel}
\def\qperp {q_\perp}
\def\vk {V_{\bf  K}}
\def\vol {\Omega}
\def\boxnorm {\frac{1}{\sqrt{\Omega}}}
\begin{document}
\input epsf
%\draft
\preprint{NORDITA-96/46 A/S/N, DOE/ER/40427-13-N96, astro-ph/9607065
}
\title{
Effects of Electron Band Structure on \\
Neutrino Pair Bremsstrahlung in Neutron Star Crusts
}
\author{
C. J. Pethick $^{1,2,3}$ and Vesteinn Thorsson$^{1,4}$ \\
}
\address{
$^{1}$ NORDITA, Blegdamsvej 17, DK-2100 Copenhagen \O, Denmark \\
$^{2}$ Department of Physics, University of Illinois at Urbana-Champaign,\\
1110 West Green St., Urbana, Illinois 61801-3080\\
$^3$ Institute for Nuclear Theory, University of Washington, \\
Box 351550, Seattle, Washington 98195-1550\\
$^4$ Department of Physics, University of Washington, \\
Box 351560, Seattle, Washington 98195-1560\\}

\date{\today}
\maketitle
\begin{abstract}
We calculate the rate of energy emission by bremsstrahlung of neutrino
pairs by electrons moving in the crystalline lattice of ions in dense
matter in the crust of a neutron star.  Since the periodic potential
in the solid gives rise to electronic band gaps which can be
as large as about 1 MeV, it is necessary in
estimating the bremsstrahlung rate at low temperatures
to take into account band structure
in detail.  We find that, in the densest parts of the inner crust of a neutron
star, neutrino emission at
temperatures of about $2\times 10^9$K or less is much suppressed
compared with earlier estimates
that treated the electron-lattice interaction perturbatively, and
conclude that neutrino pair bremsstrahlung by electrons in the crusts
of neutron stars is much less important for neutron star thermal
evolution than was previously thought.

\end{abstract}

%%%%%%%%%%%%%%%%%%%%%%%%%%%%%%%%%%%%%%%%%%%%%%%%%%%%%%%%%%%%%%%%%%%%%%%%%%%
%
%       I. Introduction
%
%%%%%%%%%%%%%%%%%%%%%%%%%%%%%%%%%%%%%%%%%%%%%%%%%%%%%%%%%%%%%%%%%%%%%%%%%%%
\section{Introduction}

The possible importance of neutrino pairs generated in collisions of
electrons with nuclei was pointed out by Pontecorvo\cite{pontecorvo},
and the first detailed calculations of the rate of the process were
made by Gandel'man and Pineev\cite{ganpin}, who considered the case of
non-degenerate, non-relativistic electrons scattering from
uncorrelated nuclei.  This work was extended by Festa and
Ruderman\cite{fesrud}, who performed calculations for degenerate
electrons.  They allowed for relativistic effects, and took into
account correlations between ions and screening of the ionic Coulomb
potential by electrons.  Their results were extended by Cazzola, De
Zotti and Saggion\cite{cds}, who performed calculations for
uncorrelated ions for a range of conditions. A general many-body
formulation of the problem of emission of neutrino pairs by electrons
scattering from ions was made by Flowers\cite{flowers}, who applied
his results to estimate bremsstrahlung by scattering from the static
lattice and also gave expressions for rates of processes in which
lattice phonons are emitted or absorbed. Flowers also stressed the
importance of the finite nuclear size in reducing the bremsstrahlung
rate. The effects of weak neutral currents was investigated by Dicus,
Kolb, Schramm, and Tubbs\cite{dicusetal}.  Further studies were made
by Soyeur and Brown\cite{soybro}, who compared energy loss by the
bremsstrahlung process with other energy loss mechanisms for various
neutron star models.  In recent years the process has been studied in
detail by Itoh and collaborators\cite{itoh}.

Basic to all the above works is the assumption that the effects of the
electron-ion interaction may be treated in lowest-order
perturbation theory, and the two perturbation theory diagrams
considered are shown in Fig.  1.  Since the energy and momentum
carried away by the neutrino pair are of order $k_B T$ and $k_B T/c$
respectively, and since the final particles are on mass shell, it is
easy to see that the intermediate energy denominator must be of order
$k_B T$.  The dimensionless parameter in the perturbation expansion is
thus of order $V_q /k_B T$, where $V_q$ is the Fourier transform of
the potential due to the ions.  For a random collection of ions, the
Fourier transform of the potential experienced by an electron will be
of order $n_Z^\frac12 {\cal V}_q / \Omega^\frac12$ where ${\cal V}_q$
is the Fourier transform of the potential due to a single ion, $n_Z$
is the number density of ions, and $\Omega$ is the volume of the
system.  Thus one does not expect problems with the perturbation
expansion in this case.  However, the situation is very different if
matter is a solid, with ions ordered on a periodic lattice. The
Fourier transform of the electron-ion interaction is then of order
$n_Z {\cal V}_K $, where $K$ is a reciprocal lattice vector and this
will lead to energy splittings in the electron spectrum of this order
of magnitude.  When the temperature is low compared with this
splitting, perturbation theory fails.  As we shall argue in detail
below, the splittings can be as large as 1 MeV in the inner crust of
neutron stars, and therefore at temperatures below about $10^{10}$ K
the first-order perturbation theory calculations of the rate fail.

In this paper we calculate bremsstrahlung of neutrino pairs by electrons
moving in the periodic potential of the ions.  Our primary purpose in this
paper is to examine {\it elastic} scattering of electrons from the lattice,
since this process was previously thought to be more important at low
temperatures than processes in which lattice phonons are emitted or
absorbed\cite{flowers}. Instead of using perturbation theory to treat the
electron-ion interaction, we first calculate band electron states, thereby
implicitly taking into account the electron-ion interaction to all orders in
perturbation theory, and then evaluate the rate of emission of neutrino pairs
when an electron makes a transition from one band state to another.  What we
find is that bremsstrahlung of neutrino pairs is suppressed exponentially at
temperatures low compared with the strength of the ionic potential, which
determines the splitting between bands.  A brief account of part of
this work has been given previously\cite{paper1}.

The paper is organized as follows.  In Sec.II we describe the basic physics
 and derive an expression
for the neutrino bremsstrahlung emissivity. Analytical expressions for the
emissivity in the high- and low-temperature limits are derived
and numerical calculations are made for intermediate temperatures in Sec.III.
Section IV contains a discussion of the results, and a comparison of neutrino
emission by bremsstrahlung from the periodic lattice with that from
conversion of phonons into neutrino pairs.

%%%%%%%%%%%%%%%%%%%%%%%%%%%%%%%%%%%%%%%%%%%%%%%%%%%%%%%%%%%%%%%%%%%%%%%%%%
%
%       II. Basic Calculations
%
%%%%%%%%%%%%%%%%%%%%%%%%%%%%%%%%%%%%%%%%%%%%%%%%%%%%%%%%%%%%%%%%%%%%%%%%%%%
\section{Basic calculations}

The basic process we consider is shown in Fig. 2, where the double
lines represent a band electron.  If the band electron propagators in
Fig. 2 are expanded to first order in the electron-ion interaction,
one recovers the two diagrams shown in
Fig. 1. To lowest order in the weak interaction, the rate for an
electron in state $1$ to emit a neutrino pair in a transition to a
state $2$, assumed to be unoccupied initially, may be found from
Fermi's Golden Rule, and is given by

\be
d{\cal R} = \frac{2\pi}{\hbar}
\delta( E_f - E_i ) |H_{fi}|^2 d\Gamma  ,
\label{golden}
\ee

\ni
where $H_{fi}$ is the matrix element of the weak interaction between the
initial electron state $i$ and the final state $f$.  The energies of all
particles in the initial and final states are denoted by $E_i$ and $E_f$.  We
denote the energies of the initial and final electrons by $E_1$ and $E_2$
%(E_1/c, {\bf p}_1)$ and $p_2 = (E_2/c, {\bf p}_2)$
respectively, and those of
the outgoing neutrino and antineutrino by $E_\nu$ and $E_{\bar \nu}$, and
$d\Gamma$ is an element of the volume of phase space.
%$q_1 = (E_\nu/c, {\bf q}_1)$ and $q_2 = (E_{\bar \nu }/c, {\bf q}_2)$.
The total rate of energy loss in
neutrino pairs per unit volume is thus given by

\be
\dot{E} =
\frac{2 \pi}{\hbar \Omega}
\sum
%\sum_{ {\bf q}_1, {\bf q}_2, {\bf p}_1, {\bf p}_2  }
%\sum_{
%       \begin{array}{c}
%       \small{{\bf q}_1, {\bf q}_2 }\\
%       \small{{\bf p}_1, {\bf p}_2 }
%        \end{array}                      }
\delta( E_f - E_i )
f(E_1) ( 1 - f(E_2) )  | H_{fi} |^2 ( E_{\nu} + E_{\bar \nu} ).
\label{edot0}
\ee

\ni
Here $f(E)$ is the Fermi function, and the factors $f(E_1)$ and $1-f(E_2)$
in Eq. (\ref{edot0}) take into account the Pauli principle, since the
transition can occur only if the initial electron state is occupied, and
the final one empty. There are no blocking factors for the neutrinos, since we
assume that neutrinos are able to escape easily, a good assumption for neutron
stars except for some seconds immediately following the birth of the star.

To calculate the rate at which an electron in one band state makes a
transition to another band state we need to evaluate matrix elements
of the weak interaction Lagrangian

\be
{\cal L} = -\sqrt{2} G {\bar \Psi}_\nu \Gamma_L^\alpha  \Psi_\nu
{\bar \Psi} ( C_L \Gamma_L^\alpha   + C_R \Gamma_R^\alpha  ) \Psi,
\label{weakint}
\ee

\ni using band states for the electrons, rather than
the plane waves of most interest in the majority of applications.
In Eq.(\ref{weakint}), $G$ is the Fermi coupling constant,
$\Gamma_{L,R}^\alpha = \gamma^\alpha (1 \mp \gamma_5 )/2$,
and in terms of the weak mixing angle $\theta_W$,
$C_L = 1 + 2\sin^2 \theta_W$ and $C_R = 2\sin^2 \theta_W $
for the emission of electron neutrinos.
The corresponding couplings for the emission of
muon and $\tau$ neutrinos,
are
$C_L'= - 1 + 2\sin^2 \theta_W$ and $C_R'= 2\sin^2 \theta_W $.

For the moment, let us be quite general, and consider transitions in which the
initial electron state is specified by the wavenumber, $\bf p$, a band index,
n, and a spin index, $\sigma$, and the final state is specified by similar
quantities with primes.
The squared modulus of the matrix element, summed over spins of the
initial and final electron, neutrino and antineutrino states,
is given by

\be
\sum_{\sigma_\nu,\sigma_{\bar \nu}} |H_{fi}|^2
= \frac{1}{\vol} 2 G^2  {\cal R}_{\alpha \beta}
\langle {\bf p}_1,\sigma_1,n_1
|J_{{\bf q}_1+{\bf q}_2}^\alpha|
{\bf p}_2,\sigma_2,n_2 \rangle
\langle {\bf p}_2,\sigma_2,n_2
|J_{-{\bf q}_1-{\bf q}_2}^\beta|
{\bf p}_1,\sigma_1,n_1 \rangle
\,\,\, ,
\label{hfisq2}
\ee

\ni where
\be
J_{\bf k}^\alpha&=& \boxnorm \int d^3x e^{-i {\bf k}\cdot{\bf x} }
                        ( C_L \Gamma_L^\alpha + C_R \Gamma_R^\alpha )
\label{current}
\,\,\, , \,\,
\ee
\ni and the neutrino part of the trace is given by

\be
{\cal R}^{\alpha \beta} &=& \sum_{\sigma_1, \sigma_2}
{\bar u}_{\sigma_1} ({\bf q}_1) \Gamma_L^\alpha v_{\sigma_2} ({\bf q}_2)
{\bar v}_{\sigma_2} ({\bf q}_2) \Gamma_L^\beta u_{\sigma_1} ({\bf q}_1)
\,\,\,.
\label{nutrace1}
\ee
%\ni
%and the matrix elements of the spatial Fourier transform of the electronic
%weak current
%are defined by
%\be
%\langle J_q \rangle
%\label{curel}
%\ee

\ni
The neutrino part may be evaluated by using the fact that for
ultrarelativistic
fermions
\be
u_{\sigma_1}({\bf k}) \otimes {\bar u}_{\sigma_2}({\bf k}) =
\frac{k\!\!\!/}{2k^0},
\label{project}
\ee

\ni and one finds

\be
{\cal R}^{\alpha \beta} =
\frac{2}{2q_1^0 2q_2^0}
( q_1^\alpha q_2^\beta + q_1^\beta q_2^\alpha - g^{\alpha \beta} q_1\cdot q_2
+ i \epsilon^{\alpha \beta \gamma \delta} (q_1)_\gamma (q_2)_\delta )
\,\,\, .
\label{nutrace2}
\ee
The electronic matrix elements depend on the neutrino momenta only through the
total four momentum of the pair, $q=(\omega/c, {\bf q})=q_1+q_2$ and
therefore one may
sum over neutrino states in the expression for the energy
emission rate, Eq.(\ref{edot0}), keeping  the total four momentum of
the pair fixed. We then use the result\cite{lenard}

\be
\frac{c(2\pi\hbar)^3}{\vol}
\sum_{{\bf q}_1,{\bf q}_2}
\delta(\omega-E_{\nu}-E_{\bar \nu})
\delta_{{\bf q},{\bf q}_1+{\bf q}_2}
\frac{q_1^\alpha  q_2^\beta}{2q_1^0 2q_2^0}
= \frac{\pi}{24}
( 2 q^\alpha  q^\beta  + g^{\alpha \beta} q^2 )  \theta( \omega - c |{\bf q}|)
\,\,\,,
\label{lenard}
\ee
to show that
\be
\sum_{{\bf q}_1,{\bf q}_2}
\delta_{{\bf q},{\bf q}_1+{\bf q}_2}
\delta(\omega-E_\nu-E_{\bar \nu})
{\cal R}^{\alpha \beta}
= \frac{\Omega}{24\pi^2\hbar^3c}
(q^\alpha q^\beta - q^2 g^{\alpha \beta}) \theta( \omega - c |{\bf q}|)
\,\,\,.
\label{nusum}
\ee
Here $\theta(x)$ is the Heaviside step function, which is unity for $x>0$
and vanishes otherwise.
The form of this result, apart from the numerical coefficient, follows
directly from the observation that the left hand side of this equation
is a second rank Lorentz tensor with vanishing divergence.
The vanishing of the divergence is a consequence of neutrino current
conservation for free neutrinos, as may be seen by contracting Eq.(6)
with $q^\alpha$.
Inserting Eq.(\ref{nusum}) into Eq.(\ref{hfisq2}), we find
\def\arraystretch{0.4}
\be
\sum |H_{fi}|^2 &=&
\frac{G^2}{12\pi^2\hbar^3c}
\sum_{
       \begin{array}{c}
       \small{{\bf q},{\bf p}_1, {\bf p}_2 }\\
       \small{\sigma_1,\sigma_2}
       \end{array}                      }
\int d\omega
\theta( \omega - c |{\bf q}|)
\nonumber\\
& &
( \langle 1 | q \cdot J_{\bf q} | 2 \rangle
\langle 2 | q \cdot J_{-\bf q} | 1 \rangle
- q^2 \langle 1 |J_{\bf q}^\alpha| 2 \rangle
\langle 2 |{J_{-\bf q}}_\alpha| 1 \rangle
) .
\label{hfisq3}
\ee
\ni
This expression is quite general: effects of band structure are included,
through the matrix elements, and no assumption has been made regarding how
relativistic the electrons are.

First we make a general remark about what transitions are possible.
The bremsstrahlung process is kinematically forbidden for initial and
final electron states in the same band, n.  The argument is
essentially identical to the one to demonstrate that (photon)
bremsstrahlung by a free electron is impossible. As a consequence of the 
fact that the
electron (group) velocity $\vec \nabla_p E$ cannot exceed $c$, the
energy difference, $E_1 - E_2$ between electron states must be less
than $cq$, where $q$ is the total momentum of the neutrino pair.
However, this energy difference, which is equal to the energy of the
neutrino pair, must exceed $cq$. Consequently, it is impossible
simultaneously to conserve energy and momentum for electron states in
the same band.  This argument does not apply to transitions between
electron states in different bands because the energy difference
between the two electron states will generally be non-zero for small
momentum transfers.

We now turn to the case of the inner crust of neutron stars, where the
electrons are highly relativistic, with Lorentz factors of 50 or
more. Under these circumstances it is an excellent approximation to
neglect the mass of the electron, since this leads to errors of order
$(m_e c^2/\mu_e)^2$, which is less than $10^{-3}$.  Here $\mu_e$ is
the electron chemical potential (neglecting electrostatic
contributions), which in the extremely relativistic limit is simply
$p_F c$, with $p_F$ being the electron Fermi momentum.  The electron
helicity is a good quantum number in this limit, and this leads to
several simplifications.  First, the left- and right-handed currents
satisfy a continuity equation $q \cdot J_{\bf q} =0$ for each
helicity, and therefore the contributions in Eq.(\ref{hfisq3})
involving such quantities vanish. The energy emission rate is thus

\be
\dot{E} &=&
- \frac{G^2 }{ 6 \pi {\Omega} \hbar^4 c }
\sum_{
       \begin{array}{c}
       \small{{\bf q},{\bf p}_1, {\bf p}_2 }\\
       \small{\sigma_1,\sigma_2}
       \end{array}                      }
\int d\omega \,\, \omega q^2
 \langle 1 |J_{\bf q}^\alpha| 2 \rangle
\langle 2 |{J_{-\bf q}}_\alpha| 1 \rangle \nonumber \\
& & \delta( E_1 - E_2 - \omega ) \,\,
\,\, f(E_1) \, ( 1 - f(E_2) ) \,\,
\theta( \omega - c |{\bf q}| ).
\label{edot1}
\ee

\ni Second,
in the term of the form $J \cdot J$ the two helicities decouple, and the net
result is proportional to $ C_L^2 + C_R^2 = 2(C_V^2 + C_A^2)$,
where $C_V=(C_L+C_R)/2$ and $C_A=(C_L-C_R)/2$.
The rate of energy emission is thus $(\frac12+\sin^2\theta_W)^2$
times the rate calculated in V-A theory, for which $C_A =C_V=1$.
The rate in muon and $\tau$ neutrinos
is obtained in a similar fashion. For them the vector and axial vector
couplings are $C'_A =1-C_A$ and $C'_V=1-C_V$\cite{dicusetal}.
The total rate of neutrino energy emission in all types of
neutrinos is thus given by replacing
$ C_V^2 + C_A^2$ with
$C_A^2 + C_V^2 + 2 ( (1-C_A)^2 + (1-C_V)^2 ) $.

The next task is to evaluate the weak interaction matrix elements, and to do
this we require a knowledge of the electronic states.  Since splittings
between bands are small compared with a typical electron energy, it is a very
good approximation for all but a few electron states to use the
almost-free-electron approximation, in which one takes into account just one
Fourier component of the periodic potential at a time.
 The component
of the lattice potential with wavenumber $\bf K$ mixes free electron states
with wavenumber $\bf k$ with those with wavenumbers $\bf k \pm \bf K$. In
the nearly-free electron approximation
(see e.g., Ashcroft and Mermin\cite{ashcroft}), one
considers only the mixing
of the two electron states that are closest in energy, and for definiteness we
denote the wavenumbers of the two states by $\bf k$ and $\bf k -\bf K$.  The
energy eigenvalues are given by

\be
E^\pm({\bf k})
= \frac{
\epsilon_{\bf k} + \epsilon_{ {\bf k}- {\bf K} }
}{2}
\pm \sqrt{
(\frac{
\epsilon_{\bf k} - \epsilon_{ {\bf k}- {\bf  K} }
}{2})^2
+ \vk^2 }
\,\,,\,\,
\label{eenergy}
\ee
\ni
where the plus sign corresponds to the upper band, and the minus sign to the
lower band.  Here $\epsilon_{\bf k}$ is the energy of a free electron, and
$\vk$ is the matrix element of the static part of the electron-ion interaction
between the two free-electron states that are mixed.  Since the helicity of
the electrons is conserved in the ultrarelativistic limit that we are
treating, the matrix element is simply the Fourier transform of the potential
for wavenumber $\bf K$ acting on an electron, multiplied by a factor giving
the overlap between the states with wavenumbers $\bf k$ and $\bf k - \bf K$,
but of the same helicity.  The helicity factor is, apart from a phase, just
$\cos(\frac 12 \theta_{\bf k , \bf k - \bf K})$, where $\theta_{\bf k , \bf k
- \bf K}$ is the angle between the two electron wave vectors.  In our
calculations the band states of greatest interest are those close to the
Fermi surface, and with a component of the wavevector along $\bf K$ close to
$\pm K/2$. For this case the modulus of the helicity factor is 
then simply $\vperp = (1
- (K/(2 k_F))^2)^\frac 12$.  The Fourier component of the electrostatic
potential due to the ions is given by $4 \pi e/K^2 \langle \rho_{\bf K}^i
\rangle$, where $ \langle \rho_{\bf K}^i \rangle$, is the statistical average
of the Fourier transform of the charge density of the ions.  As a consequence
of screening by electrons, the total electrostatic potential due to both ions
and electrons is given by the expression for ions alone, reduced by a factor
of the electronic dielectric function, $\varepsilon_{\bf K}$.  In the standard
Thomas-Fermi approximation this is given by $\varepsilon_{\bf K} = 1 +
(k_{TF}/K)^2$, where for the screening wavenumber we use the result for an
ultrarelativistic electron gas, $k_{TF}^2 =4 \alpha k_F^2/\pi$, with $\alpha =
e^2/(\hbar c)$.  In summary, the matrix elements of the lattice potential are
given by

\be
\vk = - \vperp \frac{4 \pi e \langle \rho_{\bf K}^i \rangle }{ K^2 }
= - \vperp \frac{4 \pi e^2 n_Z Z }{ K^2 }\frac{F_{\bf K} e^{-W}}
{\varepsilon_{\bf K}}
\,\,\,,\,\,\,
\label{gap}
\ee
\ni where $n_Z$ and $Z$ are, respectively, the number density and proton
number of the ions,  $F_{\bf k}$ is the form factor
reflecting the shape of the nuclear charge distribution and
and $W$ is the Debye-Waller factor describing thermal vibrations.

The evaluation of the electron trace is more complicated because of
the electron band structure.  First of all, a remark about
bookkeeping: the electron spectrum is shown in Fig. 4 as a function of
$k_{\parallel}$ the component of the electron wavenumber parallel to
$\bf K$ for a fixed value of $k_{\perp}$, the component of the
electron wavenumber perpendicular to $\bf K$.  For simplicity we take
into account only the effects of the lattice potential at a single
reciprocal lattice vector.  At low temperatures the electron states
that can participate in neutrino pair emission are those close to the
maxima and minima of the electron spectrum.  With the normal choice of
Brillouin zone boundaries, these states would lie close to the zone
boundary, and one could, for example, have processes in which an
electron on branch 1 made a transition to a state on branch 2, or to a
state on branch 4.  The fact that the wavenumbers of states on
branches 2 and 4 differ by $\bf K$ complicates the calculations, and
it is convenient to redefine the zone boundary so that the branches 3$^\prime$
and 4$^\prime$ occur in the same zone as branches 1 and 2.  All states with
$k_{\parallel} \simeq K/2$ are then given by a single expression, with
no discontinuities.  For definiteness the wavenumbers of the two plane
wave states which are strongly mixed will be denoted by $\bf k$ and
$\bf{k-K}$.

The expression for the electronic part of the matrix element may be
simplified by making use of the fact that we are interested in
processes at low temperatures, and therefore the energy and momentum
of the neutrino pair are small compared with the Fermi momentum of the
electrons, and the reciprocal lattice vectors.  In particular, the
small momentum of the neutrino pair means that the only Fourier
transforms of the current operators of interest are ones with long
wavelengths, and therefore ${\bf q}= {\bf p}_1 -{\bf p}_2$.  Matrix
elements of the current operator between the two states will possess
components with wavenumbers ${\bf p}_1 -{\bf p}_2 \pm \hbar{\bf K}$,
Umklapp terms, but these will not enter in our calculation as a
consequence of the fact that the low temperature restricts possible
neutrino momenta to values much less than $\hbar K$. We note that had
we made the conventional choice of position of the zone boundary, we
would have been forced to consider processes in which ${\bf p}_1 -{\bf
p}_2$ is comparable in magnitude to $\hbar K$.  This reflects the fact
that what are normal processes for one choice of the zone boundary may
be Umklapp processes for some other choice.  The states in the upper
and lower electron bands are

\def\arraystretch{1}
\be
\Psi^+_{{\bf p},\sigma}({\bf r}) &=& \boxnorm (
u_{\bf p} e^{\frac{i}{\hbar} {\bf p}\cdot{\bf r}} u_\sigma ({\bf p})
+ v_{\bf p} e^{\frac{i}{\hbar} ({\bf p}-\hbar {\bf K})\cdot{\bf r}}
u_\sigma ({\bf p}-\hbar {\bf K}) ) \nonumber\\
\Psi^-_{{\bf p},\sigma}({\bf r}) &=&  \boxnorm (
v_{\bf p} e^{\frac{i}{\hbar} {\bf p}\cdot{\bf r}} u_\sigma ({\bf p})
- u_{\bf p} e^{\frac{i}{\hbar} ({\bf p}-\hbar {\bf K})\cdot{\bf r}}
u_\sigma ({\bf p}-\hbar {\bf K}) )
\,\,\,.\,\,\,
\ee
The expressions for the ``coherence factors" $u_{\bf p}$ and
$v_{\bf p}$ will be given below.

In calculating long-wavelength components of the current matrix elements
it is convenient to invoke the identity

\be
{\bar u}_{\sigma_1}({\bf k}) \gamma^\mu u_{\sigma_2}({\bf k}) =
\frac{k^\mu}{k^0} \delta_{\sigma_1,\sigma_2} \,\,\,\,.
\ee

\ni We find
\be
\langle {\bf p}_2,\sigma_2,n_2|{J_{\bf q}}^\alpha |
{\bf p_1},\sigma_1,n_1 \rangle
\nonumber
\ee
\be
 = \boxnorm \frac{1}{p_F} \delta_{ {\bf q},{\bf p}_1-{\bf p}_2 }
\delta_{n_1,n_2+1}
( u_1 v_2 P^\alpha - v_1 u_2 {P'}^\alpha )
( C_L \delta_{-,\sigma_1,\sigma_2} +
 C_R \delta_{+,\sigma_1,\sigma_2 } ) + Umklapp
\,\,\,,\,\,\,
\label{elel}
\ee
where we have neglected terms of order $\Delta p/p_F$, where  $\Delta p$ is
the difference between the electron (pseudo)momentum and the Fermi momentum.
In Eq.(\ref{elel}), $u_{1,2}=u_{{\bf p}_{1,2}}$,
$v_{1,2}=v_{{\bf p}_{1,2}}$ and the electron four momenta
$p_{1,2}=(E_{1,2}/c,{\bf p}_{1,2})$
occur in the combinations
$ P = (p_1+p_2)/2 \,\,,\,\, P' = P - ( 0, \hbar {\bf K} )$.
We thus find for the electron trace
\be
\sum_{\sigma_1,\sigma_2 }
 \langle 1 |J_{\bf q}^\alpha| 2 \rangle
\langle 2 |{J_{-\bf q}}_\alpha| 1 \rangle
=
- \frac{16 \vpar^2}{\vol}
\frac{C_A^2 + C_V^2}{2}
u_1 v_1 u_2 v_2
\delta_{ {\bf q},{\bf p}_1-{\bf p}_2 } \delta_{n_1,n_2+1}
+ Umklapp
\,\,,\,\,
\label{eltrace}
\ee
\ni where $\vpar = \sqrt{1-\vperp^2}=K/(2k_F)$.
{}From Eq.(\ref{eltrace}) follows that the
energy emission rate in neutrinos is given by
\be
\dot{E} &=&
\frac{ G^2 }{ 48 \pi^7 \hbar^{10} c }
\frac{ C_A^2 + C_V^2 }{2}
\sum_{\bf K} \vpar^2
\int
d\omega d^3q d^3p_1 d^3p_2 \,\,\,\omega q^2 u_1 v_2 v_1 u_2
\nonumber \\
& &
\delta( E_1 - E_2 - \omega )\delta^{(3)} ({\bf p}_1 - {\bf p}_2 - {\bf q})
f(E_1) ( 1 - f(E_2) ) \theta( \omega - c |{\bf q}| )
\,\,\,.\,\,\,
\label{edot3}
\ee
The sum over electron bands is contained in the sum over
reciprocal lattice vectors $\bf K$ in Eq.(\ref{edot3}).
Note that since electron states were defined within
a given Brillouin zone, Eq.(\ref{edot3}) includes
a factor of 1/2 so that there is no overcounting when summing
over equal but opposite reciprocal lattice vectors.
{}From Eq.(\ref{elel}) one also finds that
$\langle 1 | q \cdot J_{\bf q} | 2 \rangle
\langle 2 | q \cdot J_{-\bf q} | 1 \rangle
\propto ( u_1 v_2 q \cdot P -  v_1 u_2  q \cdot P')^2$.
It may be verified that this vanishes
identically, as argued earlier on the basis of current conservation.

We now perform the energy and momentum integrations.
To do this we
 require expressions for the ``coherence factors'',
$u_{\bf p}$ and $v_{\bf p}$. They are identical in form to those found
in the standard treatment of nonrelativistic electrons in a weak
periodic potential and are given by
\be
u_{\bf p} = \frac{ \vk }{
\sqrt{ 2 {\cal E}_{\bf p}
( {\cal E}_{\bf p} - \xi_{\bf p} ) } }
\,\, , \,\,
v_{\bf p} = \sqrt{
\frac{ {\cal E}_{\bf p} - \xi_{\bf p} }{ 2 {\cal E}_{\bf p} }
}
\,\,\,,
\nonumber\\
u_{\bf p}^2 = \frac12 (1+\frac{ \xi_{\bf p} }{{\cal E}_{\bf p}}),\,\,\,
          v_{\bf p}^2 = \frac12 (1-\frac{ \xi_{\bf p} }
{ {\cal E}_{\bf p} }), \,\,\,
         {\rm and} \,\,\, u_{\bf p} v_{\bf p} = \frac{\vk}{2 {\cal E}_{\bf p}}
\,\,\,\,,
\label{cohfac}
\ee
\ni with  $\xi_{\bf p} = ( \epsilon_{\bf p} - \epsilon_{ {\bf
p}-\hbar{\bf K} })/2$ and  ${\cal E}_{\bf p} = \sqrt{ \xi_{\bf p}^2 +
\vk^2 }$.
We also note that,
in terms of the variables (\ref{delps}),
the energies of the upper and lower electron states are
\be
E^\pm(\dperp, \dpar ) =
 p_F c + \dperp \vperp c \pm \sqrt{ (\dpar \vpar c)^2 + \vk^2 }
\,\,\,\,,
\label{epmdel}
\ee
from which we find $\omega=c\vperp\qperp+{\cal E}_1+{\cal E}_2$.

It is convenient to express momenta in
cylindrical coordinates, with respect to $\bf K$.
Also, since the electrons participating in neutrino pair
emission are close the maxima and minima of the electron
spectrum, we define the shifted variables
\be
\dpar = p^\parallel - \vpar p_F  \,\,\, & {\rm and} & \,\,\,
\dperp = p^\perp - \vperp p_F   \,\,\,\,.
\label{delps}
\ee
We also have
\be
\qperp &=& p^\perp_1 - p^\perp_2 = \dperp_1 - \dperp_2 \nonumber \\
\qpar  &=&  p^\parallel_1 - p^\parallel_2
= \dpar_1 - \dpar_2 \nonumber \\
q_\phi &=& p_F \vperp ( \phi_1 - \phi_2 )
\,\,\,,\,\,\,
\ee
where in the last expression, we have discarded curvature terms,
which contribute only to next-to-leading order in $k_B T/p_Fc$.
The $p_2$ integration is now trivially performed, with the aid of
momentum conserving
delta function, and the $p_1$ angular integration simply gives the
constant $2 \pi$.
Since the coherence factors do not depend on $p^\perp$, the integration over
$p^\perp_1$ is straightforward and yields the result

\be
\int p_1^\perp dp_1^\perp f(E_1)(1-f(E_2)) &=&
\vperp p_F \int d\dperp_1
\frac{  f(E_1-\omega)-f(E_1) }
{ e^{\beta \omega} - 1}  \nonumber\\
&=& p_F \frac{\omega}{c} \frac{1}{e^{\beta \omega}-1}    \,\,\,,
\ee
where we have again utilized the fact that characteristic temperatures are
much smaller than the Fermi energy.
Performing the remaining energy integration finally yields
the expression for the energy emission rate per unit volume,
\be
\dot{E} =
\frac{ G^2 }{ 24 \pi^6 \hbar^{10} c^3 }
\frac{ C_A^2 + C_V^2 }{2}
\mu_e
\sum_{\bf K}
\vpar^2 \,
\int d\dpar_1 \,\, d^3q \,\,
%\int dp^\parallel_1 \,\, d^3q \,\,
\omega^2  \,  q^2 \, u_1 v_1 u_2 v_2 \,
\frac{1}{e^{\beta \omega} - 1} \,
\theta( \omega - c |{\bf q}| )
\,\,\,,
\label{edotgen}
\ee
with $\omega=E_1-E_2$.

Inserting into Eq.(\ref{edotgen}) the expressions for the coherence
factors (\ref{cohfac}) and using Eq.(\ref{epmdel}) we find
the energy emission rate,
\be
\dot{E} = \frac{ G^2 }{ 96 \pi^6 \hbar^{10} c^5 }
\frac{ C_A^2 + C_V^2 }{ 2 }
\mu_e
\sum_{\bf K}
\vpar^2 \vk^2  \int dQ_\parallel dq_\parallel dq_\phi dq_\perp
\frac{\omega^2(\omega^2-c^2 {\bf q}^2)}
{ {\cal E}_1 {\cal E}_2 ( e^{\beta \omega} -1 ) }
\theta( \omega - c | {\bf q} | )
\,\,\,,
\label{edotI}
\ee
\ni with $P_\parallel=\dpar_1-\frac12 \qpar$is an average crystal momentum
variable for the initial and final electron states. The $q_\phi$ integral in
Eq.(\ref{edotI})
may now be performed, leaving an integral over the remaining momenta,
which we express in terms of
the dimensionless variables
\be
x_1 = \frac{c P_\parallel }{ k_B T } \,\,,\,\,
x_2 = \frac{c q_\parallel }{ k_B T } \,\,,{\rm and}\,\,
x_3 = \frac{c q_\perp }{ k_B T }  \,\,\,\,.
\ee
The energy emission rate may thus be written as
\be
\dot{E} = \frac{ 4 \pi G^2 }{567 \hbar^{10} c^9 }
\frac{ C_A^2 + C_V^2 }{ 2 }
\mu_e (k_B T)^6
\sum_{\bf K}
\vpar^2 \vk^2 I(\vpar, \frac{k_B T}{|\vk|})
\,\,\,\, ,
\ee
where $I$ is the dimensionless integral
\be
I(\vpar, t) = \frac{63}{8 \pi^7}
\int_0^\infty \, dx_1 \,
\int_0^\infty \, dx_2 \,
\int_{x_3^{min}}^{x_3^{max}} \, dx_3 \,
\frac{ w^2 ( w^2 - x_2^2 - x_3^2 )^\frac32 }
{ e_1 e_2 ( e^w - 1 ) }    \,\,\,.
\label{i2x3}
\ee
The integral is to be carried out over the region where $w^2 - x_2^2 - x_3^2 $
is positive.
We have defined the dimensionless energies
\be
e_{1,2} = \sqrt{ \frac{1}{t^2} + \vpar^2( x_1\pm\frac12 x_2 )^2 }
\,\,{\rm and}\,\,\,
w = \vperp x_3 + e_1 + e_2 \,\,\,,
\ee
in terms of the variable $t =k_B T/ \vk$.
The limits of the $x_3$ integration in Eq.(\ref{i2x3}) are
\be
x_3^{max,min} = \frac{\vperp}{\vpar^2} (e_1+e_2)
\pm \frac{1}{\vpar^2}\sqrt{ (e_1+e_2)^2 - \vpar^2 x_2^2 }
\,\,\,.
\ee
The limits may be put in a more symmetric form by defining the
variable $y=x_3-\vperp(e_1+e_2)/\vpar^2$, leading to
\be
I =\frac{63}{8 \pi^7} \vpar^3
\int_0^{\infty} \, dx_1 \,
\int_0^{\infty} \, dx_2 \,
\frac{1}{ e_1 e_2 }
\int_{-a}^{+a} \, dy \,
( a^2 - y^2 )^\frac32
\frac{(\vperp y + \frac{1}{\vpar^2}(e_1+e_2))^2}
{e^{\vperp y+\frac{1}{\vpar^2}(e_1+e_2)}-1}
\,\,\,,
\label{i2y}
\ee
with $a=\frac{1}{\vpar^2}\sqrt{ (e_1+e_2)^2 - \vpar^2 x_2^2 }$.

%%%%%%%%%%%%%%%%%%%%%%%%%%%%%%%%%%%%%%%%%%%%%%%%%
%
%  III. Evaluation of the Neutrino Emission Rate
%
%%%%%%%%%%%%%%%%%%%%%%%%%%%%%%%%%%%%%%%%%%%%%%%%%%
\section{Evaluation of the Neutrino Emission Rate}
In this section we evaluate our basic result for a number of different
situations, beginning with the low- and high-temperature limits, and then we
describe results for more general cases.

\subsection{High-and Low-Temperature Limits}
Analytical results may be obtained in limiting cases.
We first consider temperatures small compared with $|\vk|$,
where our results differ dramatically from earlier ones.
This limit corresponds to $t \ll 1$.
In this case, the denominator in Eq.(\ref{i2x3})is essentially $e^w$,
and we simply expand the integrand around the state of lowest
``energy" $w$, given by
\be
x_1=x_2=0 \,\,{\rm and}\,\,x_3=x_3^{min} = - \frac{1}{1+\vperp} \frac2t
\,\,\,\,.
\ee
All three integrals are thus reduced to gaussian ones, with the result
\be
I =
\frac{ 189 }{2\pi^\frac{11}{2}
 (1-\vperp)^\frac12 (1+\vperp)^2 \vperp^\frac52 }
t^{-\frac52}
e^{x_3^{min}}
\,\,\,\,.
\label{I2lowtemp}
\ee
The rate of energy emission is therefore
\be
{\dot E}_<
 = \frac{2 G^2}{ 3 \pi^\frac92 \hbar^{10} c^9 }
\frac{ C_A^2 + C_V^2 }{2}
\mu_e
(k_BT)^\frac72 \,\,
\sum_{\bf K}
\frac{(1-\vperp)^{\frac 12} }{
\vperp^{\frac 52}  ( 1 + \vperp)   }
|\vk|^\frac92
e^{-\frac{1}{k_BT} \frac{2|\vk|}{1+v\perp} }  \,,\,  k_B T \ll |\vk|
\,.
\label{lowtemp}
\ee
\ni
The exponential dependence reflects the fact that the minimum energy of
the neutrino pair is $2|\vk|/(1+v_{\perp})$.  This is easily seen by observing
that the energy of the neutrino and antineutrino is given by $\omega = E_1 -
E_2 \ge 2|\vk| + q_{\perp}v_{\perp}$, and that, in addition, the four-momentum
of the neutrino pair must be time-like, $\omega \ge cq \ge c|q_{\perp}|$.

We next consider the limit of temperatures much greater than $\vk$,
corresponding to $t \gg 1$.  In this limit, the range of $y$ integration
in Eq.(\ref{i2y}) goes
to zero in the region $2 x_1 < x_2$.
It is furthermore convenient to perform the $x_1$ integration last,
since for fixed $x_1$, $x_2$ and $y$ lie on a disk of radius $2 x_1$.
We write
\be
I = \frac{63 \vpar}{2\pi^7}
\int_0^\infty \, dx_1 \,
\int_{-2 x_1}^{2 x_1} \, dy \,
\frac{ (\vperp y + 2 x_1)^2 }{ e^{\vperp + 2 x_1} - 1}
I_1   \,\,\,\,,
\ee
where
\be
I_1 &=& \int_0^{\sqrt{4 x_1^2 - y^2} } \, dx_2 \,
\frac{ ( 4 x_1^2 - x_2^2 - y^2 )^\frac32 }{ 4 x_1^2 - x_2^2 }
\nonumber \\
&=& \frac{\pi}{4}
( 4 x_1^2 - 3 y^2 + \frac{ |y|^3 }{ x_1 } )  \,\,\,.
\ee
The remaining integrals are performed by defining the variables
$\eta_1 = 2 x_1 + \vperp y$, associated with energy, and
the orthogonal variable $\eta_2=-\vperp x_1 + 2 y$.
The $\eta_2$ integral is elementary, and the final, $\eta_1$, integral
is evaluated using standard techniques for Bose integrals
%with limits
%$\eta_2^{min} = -\eta_1(4+\vperp)/ (2(1-\vperp))$,
%$\eta_2^{max} = \eta_1(4-\vperp)/(2(1+\vperp))$,
\be
\int_0^\infty \, d\eta_1 \,
\frac{ \eta_1^5 }{ e^{\eta_1}-1 }
= \frac{8 \pi^6}{ 63 }       \,\,\,\,.
\ee
We are thus led to the following result for the basic phase
space integral
\be
I =
\frac{1}{\vperp^2 \vpar }
( 1 - \frac{\vpar^2}{\vperp^2} \log{ \frac{1}{\vpar^2} } )
\,\,\,\, ,
\label{I2hightemp}
\ee
with the corresponding energy emission rate given by
\be
{\dot E}_> =
\frac{4 \pi G^2 }{567 \hbar^{10} c^9 }
\frac{C_A^2+C_V^2}{2} \mu_e (k_B T)^6
\,\,\sum_{\bf K} \frac{\vpar}{\vperp^2} ( 1 -
\frac{\vpar^2}{\vperp^2} \log{ \frac{1}{\vpar^2} } ) |\vk|^2
\,,\, k_B T \gg |\vk|
\,\,\,\,.
\label{hightemp}
\ee
We have checked that Eq.(\ref{hightemp}) is consistent with earlier
calculations in which the electron-lattice interaction was
treated perturbatively ( Fig.1 ).   Flowers\cite{flowers} has given
the energy emission in integral form, valid
for nonvanishing electron mass.
By evaluating these integrals
in the limit of vanishing electron mass we were able to arrive at
the analytical result (\ref{hightemp}).

\subsection{Results for Arbitrary Temperatures}

The general expression for the neutrino emission rate is rather
complicated, so we begin by discussing a number of its ingredients.

In our discussions we shall focus on the highest density regions in a
neutron star crust since most of the mass of the crust is at densities
close to that at the inner boundary of the crust, and consequently
such matter is expected to be responsible for most of the neutrino
pairs produced by electron bremsstrahlung.  An important quantity is
the strength of the periodic potential, given by Eq.(\ref{gap}).  First
of all, we note that the effects of electron screening are negligible,
since for the smallest reciprocal lattice vector for a bcc lattice,
$K=2 \sqrt 2 \pi/a$, and therefore the dielectric function is
$\varepsilon = 1 + (9 /(2 \pi^5))^{1/3} Z^{2/3}\alpha \approx 1 +
1.79\times 10^{-3} Z^{2/3}$, which differs from unity by less than
3\% for the largest Z's encountered in neutron star crusts. Next, we
turn to the nuclear form factors. At the highest densities, their
effects are significant, as we now show.  According to the
calculations of Ref.\cite{lrp}, spherical nuclei have a radius of $r_N
=8.7$ fm when spherical nuclei become unstable with respect to
formation of rod-like nuclei. The interface between nuclear matter and
the neutron liquid outside nuclei is rather diffuse, so we assume that
the nuclear charge distribution may be approximated by a gaussian
$exp(-3r^2/(2 r_{rms}^2))$, and for $r_{rms}$ we take the value
$\sqrt{3/5}r_N$, corresponding to a liquid drop with uniform proton
density.  The charge density in coordinate space is thus proportional
to $exp(-5 r^2/(2 r_N^2))$, and the form factor is given by $F_q
=exp(-q^2r_N^2/10)$.  For matter at the highest density at which
nuclei are roughly spherical, for which $\mu_e=78$MeV and
$Z=48$\cite{lrp}\footnote{We draw attention to the fact that in
Ref.\cite{lrp} the cell radius at the phase transition between
spherical nuclei and rod-shaped ones the cell radius is given as 19.2
fm.  This is a misprint, and the actual value is 17.6 fm. In our
earlier work\cite{paper1} we used the spurious value, which led to a Z
of 62.} , the form factor is 0.675 for the smallest reciprocal lattice
vector, and 0.009 for the largest one, approximately $2p_F$.

 Finally, we consider the Debye-Waller factor.  This is more difficult
to estimate since, to the best of our knowledge, there exists no
treatment of lattice vibrations in dense matter that takes into
account the effects of the neutrons between nuclei.  To make estimates
we shall use the results for the pure Coulomb lattice given by
Yakovlev and Kaminker\cite{yk}.  Again we assume conditions
appropriate for the highest density at which spherical nuclei
survive. The total number of nucleons per unit cell is $A' \approx
1466$, and the number of nucleons within the nucleus is $A \approx
427$.  Thus the total number of nucleons per proton, $x$ is
0.033. (Note that, because there are neutrons outside nuclei, $x \neq
Z/A$.) The total mass density is $1.07\times 10^{14}$g cm$^{-3}$.  The
Debye-Waller factor is

\be
W(q) \approx 0.078\frac{50}{Z}(\frac{x}{0.1})^{1/6}
(\frac{Z}{0.1A})^{1/2}\rho_{14}^{1/3}
\left(1.4e^{-9.1 t} + 12.995 t\right) (\frac{q}{2p_F})^2  \,\,.\,\,
\ee
Here $t=T/T_p$, where the ``plasma" temperature is given by
\be
T_p =\frac{\hbar \omega_p}{k_B} \approx \frac{\hbar}{k_B}
(\frac{4\pi Z^2 e^2 n_Z}{m_{ion}})^{1/2} \approx 7.8
\times 10^{9}\left(\rho_{14}\frac{x}{0.1}\frac{Z}{0.1 A}\right)^{1/2}{\rm K}
\,\,\,,\,\,\,
\label{tp}
\ee
where $\omega_p$ is the plasma frequency for the ions alone,
neglecting the presence of the outside neutrons. The quantity
$\rho_{14}$ denotes the mass density in units of $10^{14}$ g
cm$^{-3}$.  For a temperature of $10^9$ K, $e^{-W(q)}$ is 0.98 for the
smallest reciprocal lattice vector, and 0.81 for the largest one.  At
$10^{10}$ K, the corresponding values are 0.83 and 0.14 .  This
demonstrates that the effects of Debye-Waller factors can be
significant for temperatures close to the melting temperature, which
is about $1.2\times 10^{10}$ K, but the effects of the periodic
potential are still substantial for the lowest reciprocal lattice
vectors.  We emphasize that these estimates are based on the
assumption that the lattice dynamics is that of a pure Coulomb
crystal, and more realistic estimates can be made only after a
thorough investigation of lattice dynamics that allows for the
neutrons between nuclei.

For conditions under which nuclear form factors and
Debye-Waller factors may be neglected, the dependence of emission rate on
the $\mu_e$, $T$,
and Z is of the general form
\be
\dot E \sim T^6 \mu_e^3  g(Z, T/\mu_e)
\,\,\,,\,\,\,
\ee
where $g$ approaches a constant for temperatures large compared with
$Z^{2/3}\alpha \mu_e$, the magnitude of the largest band gaps.

Before presenting the results of detailed calculations we describe
qualitative features of the emission rate.  As one can see from the
asymptotic low- and high-temperature expressions Eqs.(\ref{lowtemp})
and (\ref{hightemp}), the crossover between the two sorts of behavior
occurs at a temperature $\sim 2|\vk|/(1+\vperp)/k_B$.  Thus the
crossover temperature for large wavevectors,$\sim 2p_F$, is a factor
$\sim Z^{2/3}$ smaller than for the smallest reciprocal lattice
vectors, and thus there is a large temperature range in which neither
of the limiting expressions is a good approximation.  We note that
reciprocal lattice vectors close to $2p_F$ give very small
contributions because helicity conservation ensures that electrons
cannot be scattered by ions for $K =2p_F$, as reflected by the
$\vperp$ factor in the expression for the ionic potential,
Eq.(\ref{gap}).  At high temperatures, a small reciprocal lattice
vector gives a larger contribution than a large one, because the
emission rate is proportional to the square of the Coulomb matrix
element.  As the temperature is lowered, the first reciprocal lattice
vectors for which bremsstrahlung is suppressed are the smallest ones,
because they give rise to the largest band gaps.  Consequently, at
lower temperatures, the reciprocal lattice vectors that give the
largest contributions become larger and larger, until in the
low-temperature limit, the dominant ones have wavevectors close to
$2p_F$.

We now present results of a numerical evaluation of the
the rate of energy emission from neutrino
pair bremsstrahlung.
We consider matter at the highest density at which nuclei are
roughly spherical, for which $\mu_e=78$MeV and $Z=48$,
according to the recent calculations of Ref.\cite{lrp}.
We shall neglect Debye-Waller factors.

In Fig. 5 we plot
energy emission rates, for several values of
$\vpar = K/(2k_F)$, compared with the corresponding
quantities in the high temperature limit.
The solid lines in Fig.5 are for nuclear form factor $F_q=1$,
and the dashed lines are for $F_q=\exp(-q^2r_N^2/10)$, where
$r_N=8.7$fm.

The total energy emission rate is obtained by evaluating the basic phase
space integral, Eq.(\ref{i2y}) for all reciprocal lattice vectors
entering in the sum of Eq.(\ref{edotI}).
The relevant reciprocal lattice vectors are those of length $K < 2 k_F$,
where $k_F$ is the electron Fermi momentum.
The Fermi sphere is of radius
$R=(3 Z / (32 \pi))^{1/3}$ in
units of $4 \pi/ a$,  the side length of the conventional cubic cell
of the reciprocal ( fcc ) lattice.  Here,
$a=2/n_Z$ is the length of a side in the conventional cubic
cell of the direct ( bcc ) lattice.
For $Z=48$ we find that $R=1.13$, and that  there are 200 reciprocal lattice
vectors contributing to the sum.

The numerical evaluation may be streamlined
by exploiting the symmetry of the lattice when evaluating
phase space integrals. This becomes especially relevant when
there is a large number of reciprocal lattice vectors
contributing to the energy emission rate.
A given reciprocal lattice vector has
47 partners of equal length, if the three vector components differ in
magnitude, and none are zero.
The degeneracy is reduced if two or more components are equal, or
if one or more components are zero.  To give an idea what is to be
gained from such considerations, we
note that the 200 reciprocal lattice vectors contributing
to neutrino emission have only 11 distinct lengths.

The three dimensional integral (\ref{i2y}) was evaluated using
the extended trapezoidal
rule\cite{numrec}.  We checked that both the low temperature (\ref{I2lowtemp})
and high temperature (\ref{I2hightemp}) limits were reproduced to within
better than 1\% for a range of values of $\vpar$, corresponding to
different choices for reciprocal lattice vectors.

In Fig.6 we show the rate of energy emission from neutrino pair
bremsstrahlung, divided by the high temperature rate,
Eq.(\ref{hightemp}), for the conditions we have described.
The solid line shows this ratio for nuclear form factor $F_q=1$.
The short dashed line shows the corresponding ratio for
$F_q=\exp(-q^2r_N^2/10)$, with $r_N=8.7$fm.
The rate from the 12 smallest reciprocal lattice
vectors alone divided by the total high temperature rate
is given by the line with alternating short and long dashes,
for $F_q=1$.
The long dashed line gives the emission
rate due to conversion of thermal phonons into neutrino pairs,
Eq.(44), as compared with Eq.(40).

%%%%%%%%%%%%%%%%%%%%%%%%%%%%%%%%%%%%%%%%%%%%%%%%%
%
%  IV. Discussion of Results
%
%%%%%%%%%%%%%%%%%%%%%%%%%%%%%%%%%%%%%%%%%%%%%%%%%%

\section{Discussion of Results}

An unexpected feature of the results is that the ratio $\dot E/T^6$ is
not a monotonic function of temperature, and it has a maximum.  We do
not believe this to be an artifact of the numerical calculations for
two reasons.  First, the numerical calculations agree with the
analytical results. This is a non-trivial check since the regions of
integration in the high- and low-temperature limits are very
different.  Second, we have performed the numerical integrations by
two different methods, with identical results.  It is interesting to
note that non-monotonic temperature dependence was predicted
theoretically and subsequently observed in a quite different physical
context, the relaxation of nuclear spins in superconductors.  We
surmise that the origin of the non-monotonic behavior in the case of
neutrino emission may be a similar effect, because the excitation
spectrum of electrons near a band gap is similar to that for the
excitations in a superconductor.  However, the neutrino emission
process we have considered is an interband process, and thus its
analog in a superconductor would be the annihilation of two
positive-energy excitations.  However, the relaxation of a nuclear
spin in a superconductor takes place by scattering of one positive
energy excitation to another positive energy state, which is analogous
to an intraband process in band theory.  It is desirable to confirm
this analytically, by, for example, calculating the leading
corrections to the high-temperature limit.

While the suppression of the bremsstrahlung rate for processes with a
particular reciprocal lattice vector is rather abrupt, the suppression of
the total rate from all reciprocal lattice vectors is less abrupt, as a
consequence of the fact that the characteristic temperatures for suppression
span a range of order $Z^{2/3}$.

{}From Fig.6 we see that the smallest reciprocal lattice vectors
give almost 2/3 of the total rate at high temperature.
As the temperature is lowered, this fraction drops rapidly,
since the emission from the smallest reciprocal lattice vectors
is the first to be suppressed.

Figs.5 and 6 also illustrate the effect of nontrivial nuclear form factors.
For a given reciprocal lattice vector $\bf K$,
the crossover temperature is lower if the nuclear
form factor $F_K$ differs from unity than
if $F_K=1$, since the matrix element of the lattice potential is smaller.
For large $K$, where $F_K$ can be very small,
this effect is much more pronounced
than for small $K$.
However, the contribution of large reciprocal lattice
vectors to the total rate is small,
due to the $\vperp$ factor in the ionic potential.
Consequently,
the magnitude of the change resulting from nontrivial
nuclear form factors is closer to that given
by the smallest reciprocal lattice vectors.

In our calculations we have assumed that the periodic potential mixes
only pairs of free electron states.  However, for points of high
symmetry in the Brillouin zone, it is possible for more that two
free-electron states to be degenerate.  At high temperatures such
points will be of little significance because the phase space for
transitions to states in their vicinities is small compared with the
total.  However, at low temperatures they could be significant if it
happened that two bands could cross, even when the effects of the
periodic potential are taken into account, since this would give the
possibility of the neutrino emissivity at low temperatures behaving
as a power of the temperature, rather than exponentially.  With this
in mind we have investigated the degeneracy of bands at points of high
symmetry.  The standard discussion of this problem for terrestrial
solids \cite{jones,cracknell} cannot be carried over to case of dense
matter because electrons are highly relativistic, and therefore
spin-orbit coupling is strong.  However, by generalizing the standard
treatment to take into account electron spin and spin-orbit coupling
by employing double groups, one can demonstrate that for the bcc
structure, bands cannot be degenerate, and therefore the neutrino
emissivity will behave exponentially at low temperatures.  We conclude
that points of high symmetry will not play an overwhelming role at low
temperatures, and consequently our calculations that allowed for
mixing of only two wavenumbers at a time will give a reliable
estimate.

Another process that can generate neutrino pairs in the crusts of
 neutron stars is the conversion of thermal phonons into pairs.  This
 has been considered by many authors\cite{flowers,itoh}, and
 most recently by Yakovlev and Kaminker\cite{yk}.  One important
 characteristic temperature is the ``plasma" temperature defined in
 Eq.(\ref{tp}), in the approximation that the effects of neutrons on
 the lattice dynamics are neglected, except insofar as they affect the
 relationship between the total mass density and the electron density.
Another important temperature is the melting temperature,
$T_m=Z^2e^2/(r_c k_B \Gamma_m) \approx 2.6\times 10^{10}(Z/60)^{5/3}
(\rho_{14}x/0.1)^{1/3}$ K,  where for the
characteristic coupling parameter at melting, $\Gamma_m$ we have taken the
value 172
obtained by Nagara et al.\cite{nagara}
The quantity $r_c=(4 \pi n_Z/3)^{-1/3}= (8\pi/3)^{1/3}a$, where the latter
result is for a bcc lattice.
For temperatures larger than about $T_p/16$, the energy emission is given
approximately by
\be
\dot E_{phonon} \approx 4.2 \times 10^4 (x/0.1)^{2/3}\rho_{14}^{-1/3}
T_9^7  {\rm erg} ~{\rm g}^{-1} ~{\rm s}^{-1} .
\ee
Thus at the melting temperature, the neutrino emission due to the phonon
process is comparable in magnitude
to that from bremsstrahlung from the static lattice.

The phonon processes, which at high temperatures are primarily
Umklapp ones, begin to be cut off at a temperature for which
the wavenumber of a thermal phonon is comparable with the range of wavenumbers
for which the electron spectrum is significantly distorted from the
free-electron form in the vicinity of a band gap.  This amounts to the
condition that the thermal energy must be less than or of order $c_s/c$ times
the band gap, where $c_s$ is a typical sound speed\cite{raikh}.  Thus
phonon-induced
bremsstrahlung will be suppressed at temperatures which are smaller by a
factor of order $c_s/c$
than the temperatures at which bremsstrahlung from a static lattice is
suppressed.  For the lowest reciprocal lattice vectors, the corresponding
temperature is of order $T_p Z^{1/3}\alpha$, while for the largest reciprocal
lattice vectors it is of order $T_p Z^{-1/3}\alpha$. At the lowest
temperatures phonon-induced bremsstrahlung of neutrinos will take place by
normal processes, as opposed to Umklapp ones, and the emission will vary as
$T^{11}$\cite{flowers}. Thus as the temperature is lowered, phonon-induced
processes
will become more important that bremsstrahlung from the static lattice.
However, as one can see from Fig. 6, where the rate for emission of
neutrino pairs by phonons is
shown, calculated from Eq.(44), the phonon rate remains below the rate for
the static lattice at temperatures above about $10^8$ K.    We remark
that if one were to use an expression  more accurate than Eq.(44) for
the rate of the phonon process, this conclusion would not be altered. In a
future paper we shall make a more detailed comparison of the two
processes\cite{kpty}.

%%%%%%%%%%%%%%%%%%%%%%%%%%%%%%%%%%%%%%%%%%%%%%%%%
%
%  V. Conclusion
%
%%%%%%%%%%%%%%%%%%%%%%%%%%%%%%%%%%%%%%%%%%%%%%%%%%

\section{Conclusion}

The most important lesson to be learned from these calculations is
that, even though the electron-ion interaction has a small effect
on the energy per electron, since it is of order $Z^{2/3}\alpha \mu_e$
per electron, it can have a major effect on kinetic process at
temperatures low compared with band gaps.  For the process we have
considered in this paper, the effects are large because they involve
interband transitions.  For processes which are mediated by intraband
transitions the effects will generally be milder because the electron
states in the vicinity of band gaps will be relatively less important.

 Central to our calculations is the nature of the electronic band structure
for ultra-relativistic electrons.  This is very similar to the
non-relativistic case treated in standard texts on condensed matter physics,
but with the important difference that for the ultra-relativistic case the
matrix elements of the potential contain the familiar helicity factor which
cuts down the amplitude for large-angle scattering.  This factor has
previously been considered in the scattering of relativistic electrons by
phonons\cite{raikh}.

    The basic conclusion of our calculations is that the rate of emission of
electron pairs by electrons is strongly suppressed at temperatures below the
band gap.  For the inner crust of neutron stars, the suppression is
substantial at temperatures of order $10^9$ K, or precisely in the temperature
range where it was earlier argued that it would be the dominant neutrino
emission process should the processes in the core be suppressed by
superfluidity of the neutrons and protons, which are estimated to have
transition temperatures that can range up to about $10^{10}$ K. Of course,
estimates of superfluid transition temperatures are uncertain, but our results
suggest that the range of temperatures where neutrino bremsstrahlung by
electrons could play a significant role in neutron star cooling are extremely
limited.

    There are a number of issues that need to be addressed in order to arrive
at better estimates of the rate of neutrino bremsstrahlung emission by
electrons.  A major one is the nature of the collective oscillations of matter
in the inner crust which enter the estimates of Debye-Waller factors.  At the
lowest densities, the density of neutrons outside nuclei is low and they have
little effect on the collective modes, which are just the vibrations of the
ionic lattice, with a neutralizing background of electrons.  At higher
densities, the role of the neutrons becomes more important, and one must
consider a lattice of nuclei coupled to a neutron fluid, a problem that has
been addressed by Sedrakian\cite{sedrakian}.  However, the problem is not a
trivial one, as may be seen from the fact that
the most straightforward estimate of the effective mass of a nucleus moving
through a normal Fermi liquid leads to a divergent answer\cite{pethick}.   The
collective mode structure of the liquid-crystal-like phases with non-spherical
nuclei is a completely open problem, but these modes will come to resemble the
corresponding giant resonances in nuclear matter as the transition to the
uniform phase is approached.

We wish to acknowledge helpful correspondence with
D. Yakovlev.  The research on which this article is based
was partially supported by U.~S. National Science Foundation
grant NSF AST 93-15133, and by NASA grant  NAGW-1583.  Part of the work
carried out while one of the authors (CJP) enjoyed the hospitality of the
Institute for Nuclear Theory at the University of Washington, and support
from the Department of Energy grant to the Institute.

\begin{center}

% Figure 1
\begin{figure}
\leavevmode
\epsfbox[0 -50 523 224]{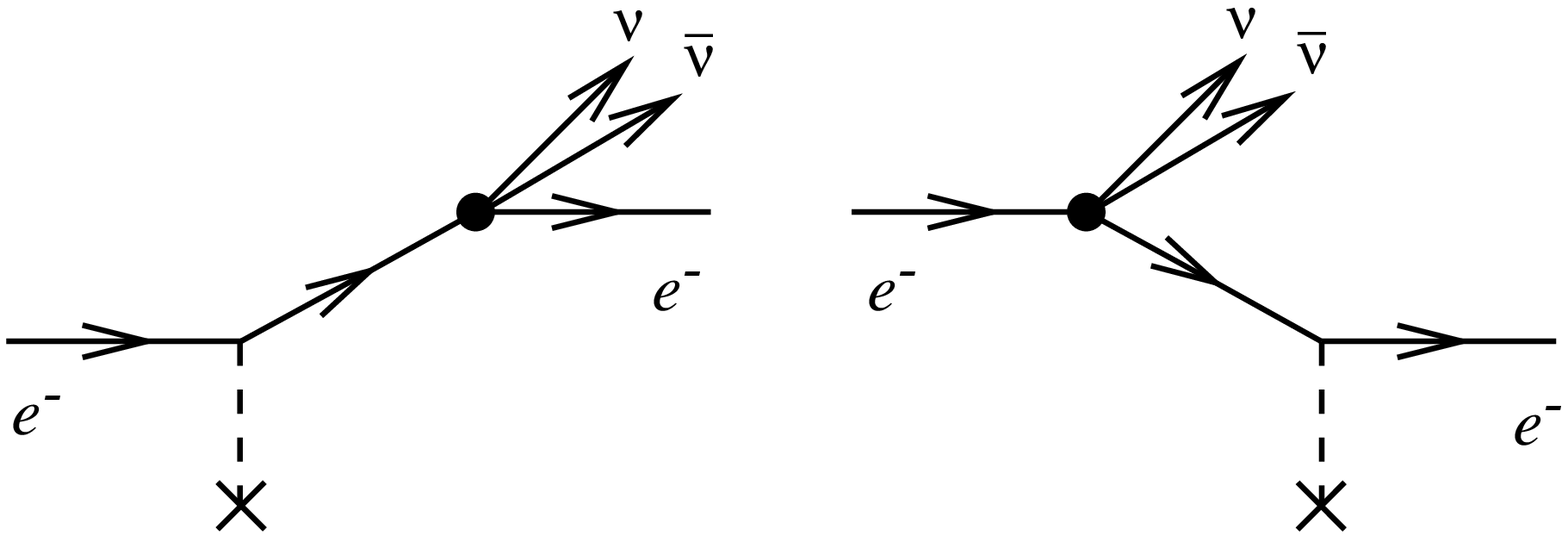}
\caption{The basic bremsstrahlung process in first-order perturbation
theory. The cross denotes an electron-lattice interaction, and the
propagators are free ones.}
\end{figure}

%Figure 2
\begin{figure}
\leavevmode
\epsfbox[0 -50 223 174]{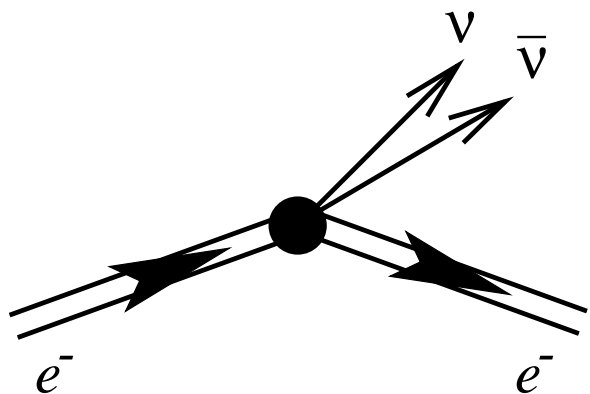}
\caption{The basic bremsstrahlung process.  The double line is the
propagator for a band electron.}
\end{figure}

%Figure 3
\begin{figure}
\leavevmode
\epsfbox[ 0 -50 401 475]{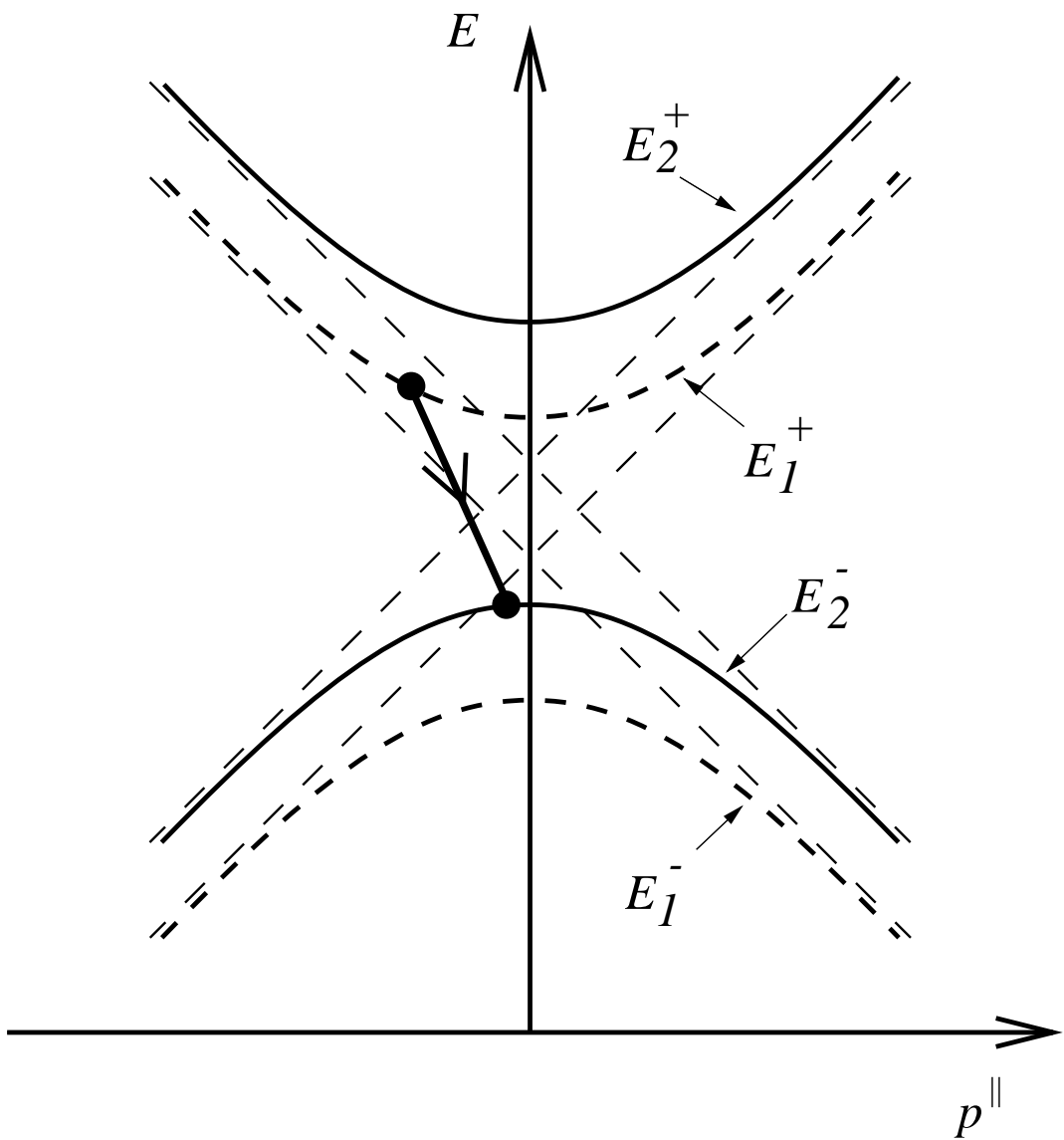}
\caption{Electron energy $E$ as a function of $p^{\parallel}$ for
two values of $p^{\perp}$. The arrow shows a possible transition.}
\end{figure}

%Figure 4
\begin{figure}
\leavevmode
\epsfbox[ 0 -50 448 450]{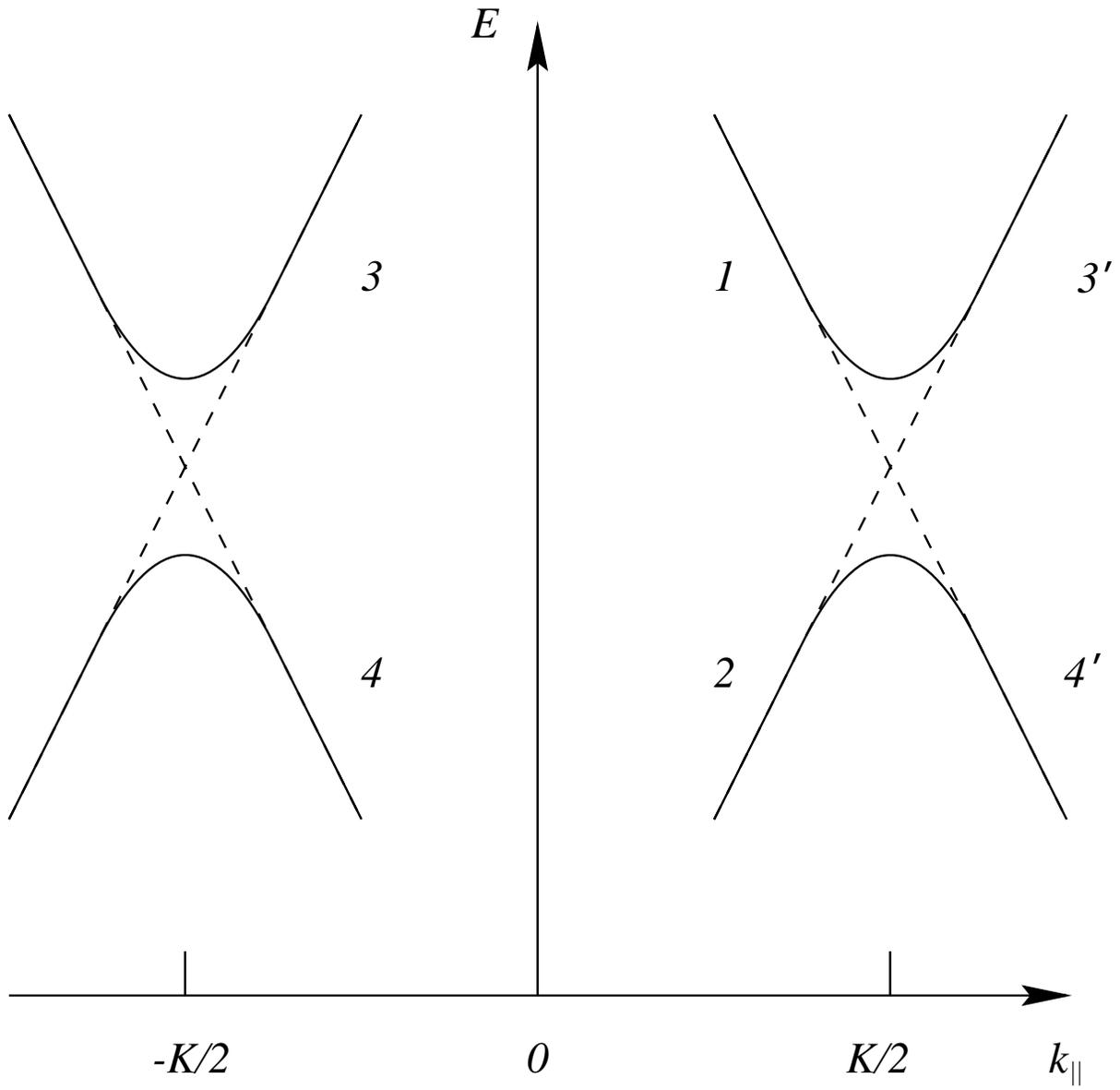}
\caption{Brillouin zone boundary. The zone boundary is defined
such that the branches of the electron spectrum labelled 1, 2,
3$^\prime$ and 4$^\prime$ occur in the same zone.}
\end{figure}

%Figure 5
\begin{figure}
\leavevmode
% Use if fig.w.caption is too high
\epsfysize=18cm
\epsfbox{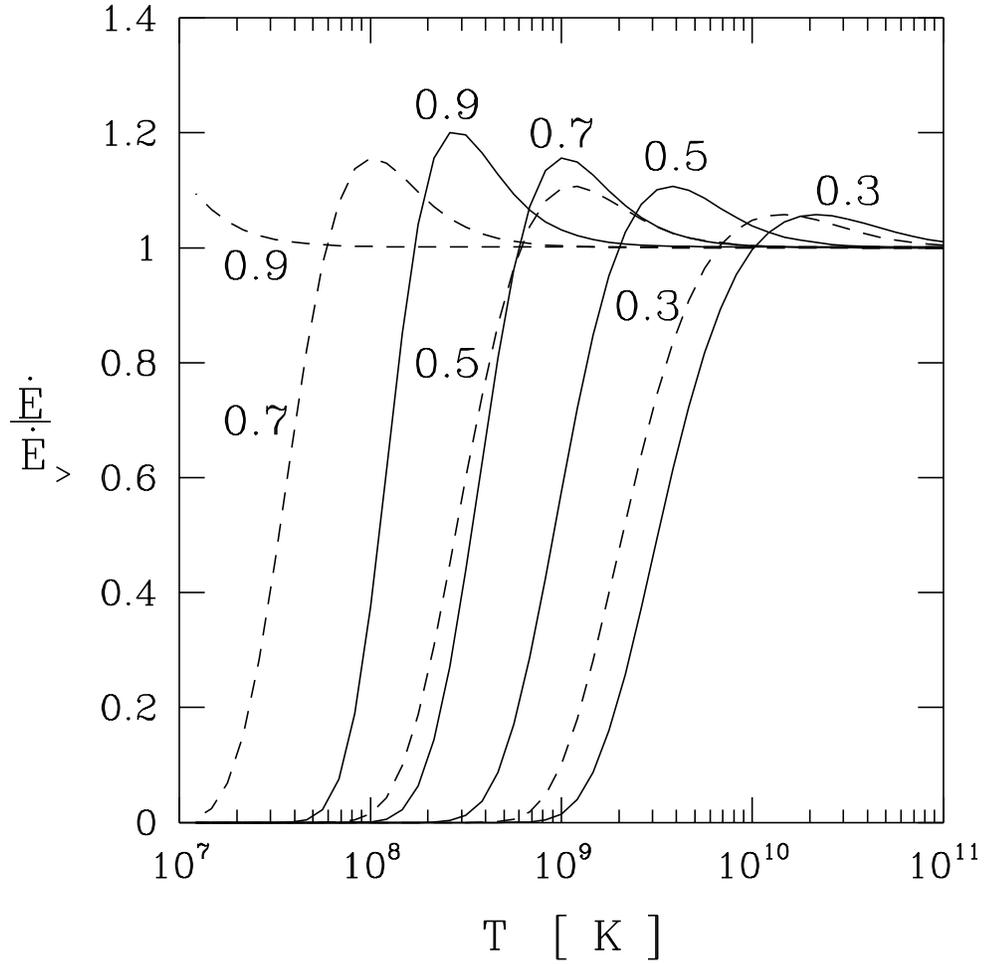}
\caption{Energy emission rates, for several values of
$\vpar = K/(2k_F)$, compared with the corresponding
quantities in the high temperature limit.
Solid lines:  Form factor $F_q=1$ ;
Dashed lines: Form factor $F_q=\exp(-q^2r_N^2/10)$,
$r_N=8.7$fm.
}
\end{figure}

%Figure 6
\begin{figure}
\leavevmode
% Use if fig.w.caption is too high
\epsfysize=18cm
\epsfbox{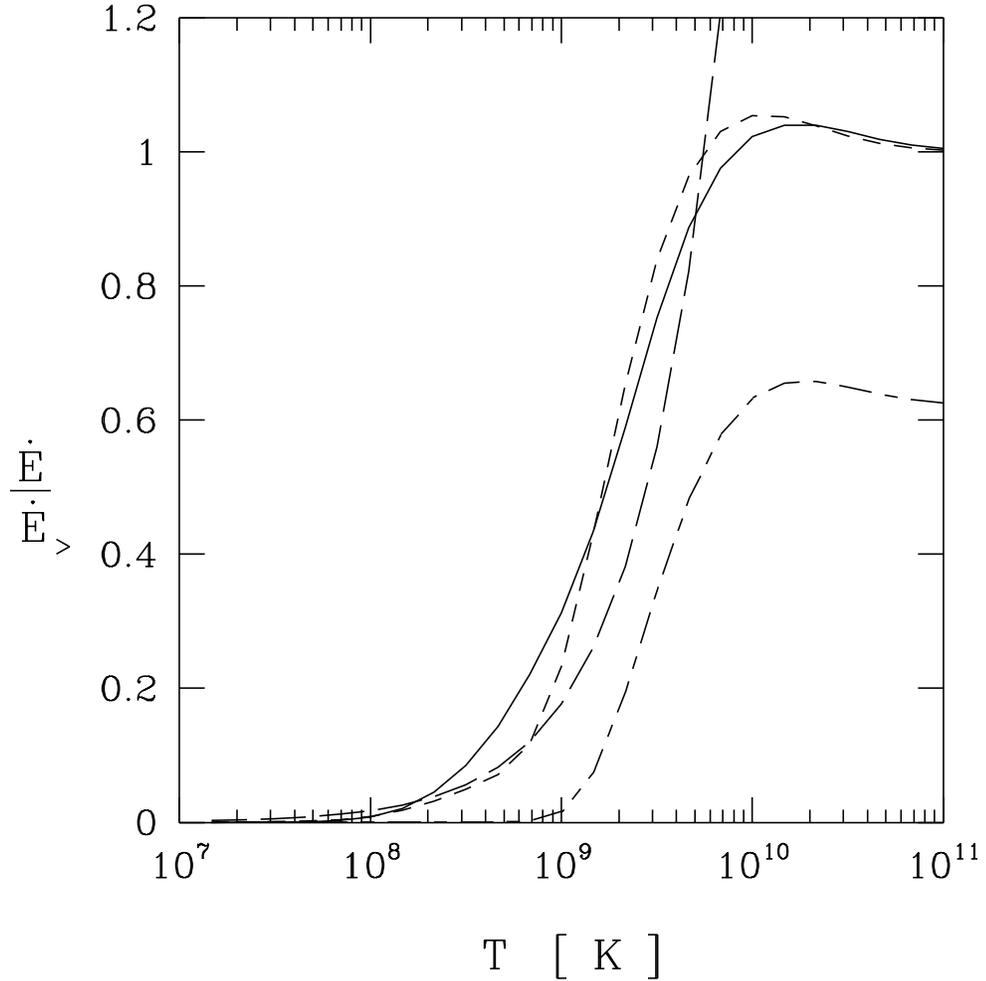}
\caption{Solid line: Energy emission rate, Eq.(28),
compared with the high temperature limit, Eq.(40),
as a function of temperature, with form factor $F_q=1$.
Alternating short and long dashes: Same as above, but
taking only the contribution of the smallest reciprocal
lattice vectors to Eq.(28).
Short dashes: Same as solid line, but with
form factor $F_q=\exp(-q^2r_N^2/10)$,
$r_N=8.7$fm.
Long dashes: The emission
rate due to conversion of thermal phonons into neutrino pairs,
Eq.(44), as compared with Eq.(40).}
\end{figure}

\end{center}

\end{document}